\documentclass[a4paper,11pt]{article}
\usepackage{pos}
\usepackage{enumitem}
\usepackage{multicol}
\usepackage{tikz}
\renewcommand{\logo}{\relax}

\title{From collider to cosmic rays: Pythia 8/Angantyr\\ for air shower simulations in CORSIKA 8}
\ShortTitle{Pythia 8/Angantyr for air shower simulations in CORSIKA 8}

\author*[a]{Chloé Gaudu}
\author[b]{Maximilian Reininghaus}
\author[c]{Felix Riehn}
\onbehalf{on behalf of the CORSIKA 8 collaboration\footnote[2]{email: \href{mailto:corsika8@kit.edu}{\texttt{corsika8@kit.edu}}}\newline\textnormal{(a complete list of authors can be found at the end of the proceeding)}}

\affiliation[a]{Bergische Universität Wuppertal, Gaußstraße 20, 42119 Wuppertal, Germany}
\affiliation[b]{Independent researcher, Germany}
\affiliation[c]{Technische Universität Dortmund, August-Schmidt-Straße 4, 44221 Dortmund, Germany}

\emailAdd{gaudu@uni-wuppertal.de}
\renewcommand{\printHeadAuthors}{Gaudu et al.}

\abstract{The simulation of extensive air showers is pivotal for advancing our understanding of high-energy cosmic ray interactions in Earth's atmosphere. The CORSIKA 8 framework is being developed as a modern, flexible, and efficient tool for simulating these interactions with a variety of high-energy hadronic models. We present the ongoing implementation and validation of Pythia 8/Angantyr within CORSIKA 8. Pythia 8, successfully used in collider physics, provides a detailed and well-tested treatment of hadronic interactions, while the Angantyr model extends its capabilities to describe heavy-ion collisions in a consistent manner. With the inclusion of Pythia 8, the CORSIKA 8 suite now enables further tuning possibilities, improving the exploration of hadronic interactions in air showers.

In this contribution, we compare the capability of Pythia 8/Angantyr to reproduce fundamental observables of high-energy particle collisions -- inelastic cross-sections and multiplicities -- to that of several established high-energy interaction models in air shower simulations. We further compare the predictions for key air shower properties, including longitudinal shower development and muon content, for iron-induced shower.}

\ConferenceLogo{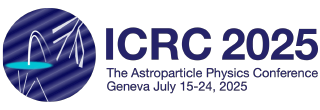}

\FullConference{39th International Cosmic Ray Conference (ICRC2025)\\
 15–24 July 2025\\
Geneva, Switzerland\\}
\begin{document}
\maketitle

\section*{Introduction}

Accurate modeling of high-energy hadronic interactions is crucial for reliable simulations of extensive air showers in astroparticle physics. CORSIKA 8~\cite{Engel:2018akg, Alameddine:2024cyd, Riehn:ICRC2025}, the modern successor to the widely used CORSIKA 7 framework, is designed with modularity and extensibility in mind, enabling seamless integration of diverse physics models. While existing high-energy interaction models like EPOS-LHC~\cite{Pierog:2013ria}, QGSJet-II.04~\cite{Ostapchenko:2010vb} and Sibyll 2.3d~\cite{Riehn:2019jet} have been widely used in air shower simulations, exploring alternative approaches is essential for understanding model-dependent uncertainties. The implementation of Pythia~8 in CORSIKA~8 enables direct comparisons with existing interaction models and provides a new tool to study hadronic physics in air showers. %This integration aims not only to 
This effort not only broadens the range of available models but also facilitates cross-disciplinary synergy between collider and cosmic ray physics. Previous studies validated the implementation of Pythia~8/Angantyr in CORSIKA~8 using vertical and inclined proton-induced air showers at energies of 10$^{17}$ and 10$^{17.5}$~eV~\cite{Gaudu:2024rsq, Gaudu:2025K1}. With this work, we extend the analysis to inclined iron-induced air showers.

\section{Pythia 8/Angantyr}
Unlike established hadronic interaction models such as Sibyll~2.3d and QGSJet-II.04, but similar to EPOS-LHC, Pythia~8~\cite{Bierlich:2022pfr} was originally designed as a general-purpose event generator, optimized for collider-based environments involving processes like e$^{+}$e$^{-}$, pp, $\overline{\mathrm{p}}$p collisions. Pythia~8 has recently been extended to allow for extensive air shower simulations. While Pythia is known for its robust description of pp collisions at the LHC, the Angantyr model~\cite{Bierlich:2018xfw, Bierlich:2021poz} extends the capabilities of Pythia to pA and AA systems by constructing these as superpositions of multiple binary sub-collisions, reminiscent of the Fritiof model~\cite{Andersson:1992iq}. Another handling of nuclear interactions using the PythiaCascade approach is possible, as discussed in~\cite{Sjostrand:2021dal}.

\vspace*{-0.4cm}
\begin{figure}[h]
  \centering
  \includegraphics[width=0.85\linewidth]{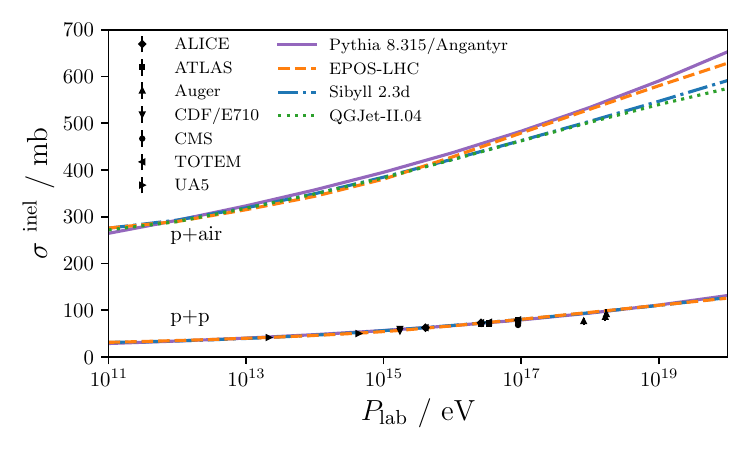}
  \vspace*{-0.5cm}
  \caption{Inelastic cross-sections as a function of the momentum in the laboratory frame of the projectile for proton-proton and proton-air collisions for Pythia 8.315/Angantyr (solid), EPOS-LHC (dashed) and QGSJet-II.04 (dotted) assuming the air composition to be 78\% $^{14}$N, 21\% $^{16}$O, 1\% $^{40}$Ar, while Sibyll~2.3d (dash-dotted) uses a simplified 80\% $^{14}$N to 20\% $^{16}$O mix. Experimental data taken from~\cite{Tkachenko:ICRC2025}.}
  \label{fig:xs}
\end{figure}

As part of our efforts to bridge collider and air shower physics, we assess how well these interaction models reproduce fundamental observables of particle production seen in high-energy collisions. In Fig.~\ref{fig:xs} a comparison of the inelastic cross-sections for proton-proton and proton-air collisions between Pythia~8/Angantyr and the most commonly used models in air shower physics is shown. The predicted proton-proton cross sections are in close agreement among the models. While EPOS-LHC and Pythia~8/Angantyr agree within 25 mb ($\sim4$\%), the difference between Pythia~8/Angantyr and QGSJet-II.04 increases to as much as 80 mb ($\sim14$\%) for the proton-air cross section. This suggests that Pythia~8/Angantyr and EPOS-LHC yield comparable interaction lengths in air showers, whereas QGSJet-II.04 is expected to produce a first interaction deeper in the atmosphere.

\begin{figure}[h]
  \hspace*{-1cm}
  \includegraphics[width=1.1\linewidth]{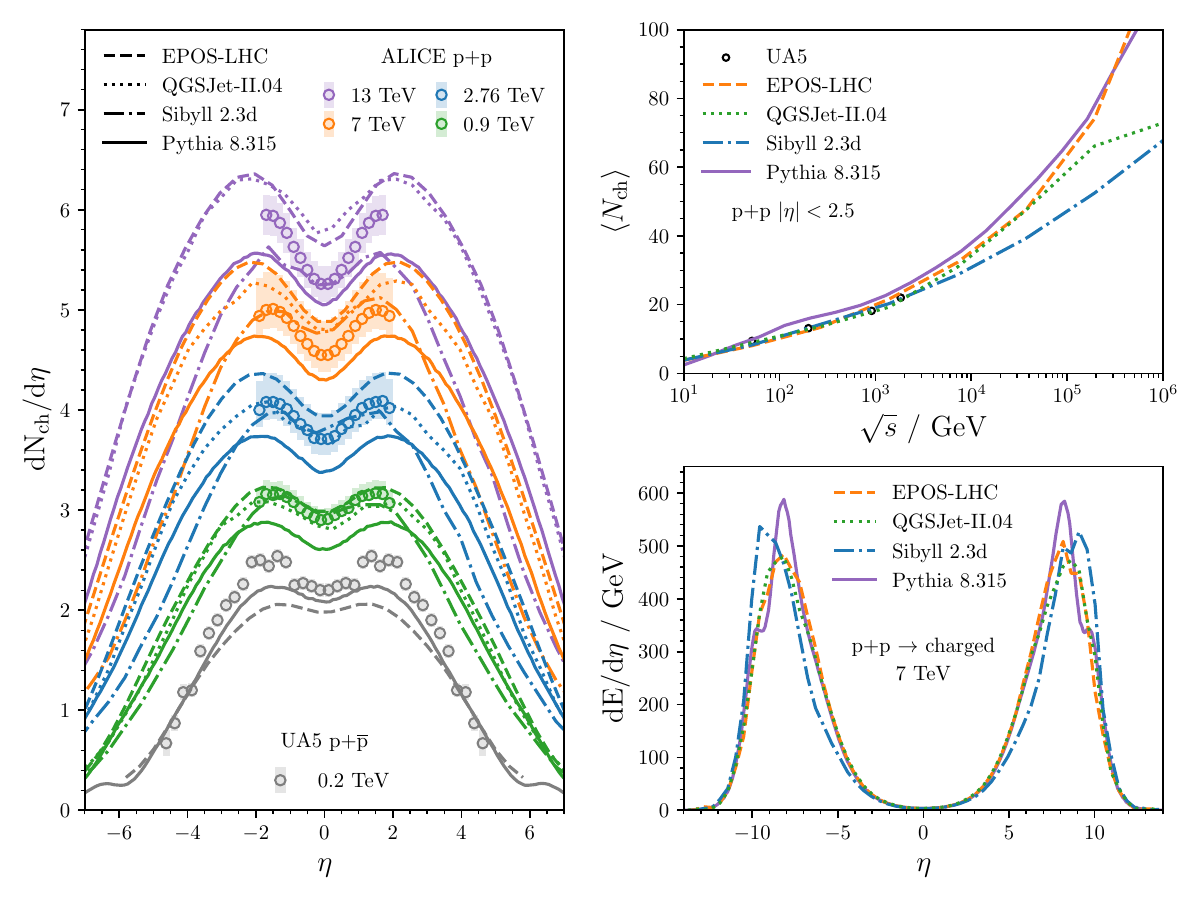}
  \vspace*{-0.7cm}
  \caption{%
  Left: dN$_\mathrm{ch}$/d$\eta$ distributions, from 0.2 to 13 TeV, in pp and p$\overline{\mathrm{p}}$ collisions. Experimental measurements are taken from ALICE~\cite{ALICE:2015qqj, ALICE:2015olq, ALICE:2010cin} and UA5~\cite{UA5:1986yef} (circle markers), and the predictions are from several interaction models: Pythia~8.315 (solid), EPOS-LHC (dashed), Sibyll~2.3d (dash-dotted) and QGSjet-II.04 (dotted). Inspired by~\cite{Basu:2020jbk}. Upper right: Charged particle multiplicity $\langle \mathrm{N}_\mathrm{ch} \rangle$ at mid-rapidity ($|\eta|<2.5$) as a function of center-of-mass energy $\sqrt{s}$ in pp collisions. Experimental measurements are taken from UA5~\cite{UA5:1986yef}. Lower right: Predictions of dE/d$\eta$ distribution in pp at 7 TeV collisions.}
  \label{fig:pp}
\end{figure}

Charged particle multiplicity at mid-rapidity ($|\eta|<2.5$), pseudorapidity density, and the energy density are essential observables in proton-proton collisions for understanding particle production. Their energy dependence, particularly the average charged particle multiplicity versus center-of-mass energy, is shown in Fig.~\ref{fig:pp}. Untuned Pythia~8/Angantyr underestimates the increase of central charged multiplicity seen by ALICE (Fig.~\ref{fig:pp}, left) and overestimates multiplicities in non-single diffractive measurements from UA5 (Fig.~\ref{fig:pp}, upper right). The energy flow, a key quantity for air shower development, shows an unusual double-peak structure (Fig.~\ref{fig:pp}, lower right). Presumably a reevaluation of the handling of projectile remnants as discussed by~\cite{Fieg:2023kld} could improve the description. 

%Proton-oxygen collisions offer an intermediate step between pp and heavy-ion interactions, relevant to the fixed-target conditions found in air showers propagating through the atmosphere. Fig.~\ref{fig:pO} displays model predictions for pO collisions, in light of the recent runs at the LHC, at $\sqrt{s_\mathrm{NN}} = 9.6$ TeV and 70.9 GeV.

%\begin{figure}[h]
%  \hspace*{-1cm}
%  \includegraphics[width=1.1\linewidth]{fig/subplot_pO.pdf}
%  \vspace*{-0.7cm}
%  \caption{%
%  Predictions of dN/d$\eta$ (left) distribution in pO collisions at $\sqrt{s_\mathrm{NN}} = 9.6$ TeV and 70.9 GeV, as well as and dE/d$\eta$ distribution at $\sqrt{s_\mathrm{NN}} = 9.6$ TeV (right) in anticipation of the recents runs at LHC, from several interaction models: Pythia~8.315/Angantyr (solid), EPOS-LHC (dashed), Sibyll~2.3d (dash-dotted) and QGSjet-II.04 (dotted).}
%  \label{fig:pO}
%\end{figure}

\section{Integration of Pythia~8/Angantyr in CORSIKA~8}
The implementation of Pythia~8 in CORSIKA~8 builds upon previous studies~\cite{Reininghaus:2022gnr, Reininghaus:2023ctx}, with a recent transition from using PythiaCascade to the Angantyr model, both available in Pythia~8.315 for the treatment of nuclear interactions. A recent feature enables single instances of Pythia and Angantyr to operate with dynamic switching of the beam system and variable collision energy on an event-by-event basis. Angantyr treats collisions with nuclear projectiles and/or targets, while Pythia is responsible for all other collision systems. The treatment of nuclear projectiles and their fragments within CORSIKA~8 has been implemented specifically for the Angantyr model. 

A comprehensive database of total and partial cross sections was compiled to support simulations of particle interactions with the atmosphere using Pythia within CORSIKA~8. The dataset includes a wide range of projectiles -- pions, kaons, protons, neutrons, other long-lived mesons and baryons, and stable isotopes from $^{2}$H to $^{56}$Fe -- as well as various targets relevant for modeling the Earth's atmosphere and comparisons with experimental data, such as protons, $^{12}$C, $^{14}$N, $^{16}$O, and $^{40}$Ar, which can be easily extended to allow for any medium.

\section{Air shower simulations}

We performed simulations of air showers induced by proton and iron primaries, with a primary energy of 10$^{19}$ eV, for an inclined geometry (zenith angle of 67$^\circ$), using four high-energy interaction models: Pythia~8.315/Angantyr, EPOS-LHC, Sibyll~2.3d and QGSJet-II.04. The particle tracking energy thresholds were set at 1 MeV for electrons, positrons, and photons, and 1 GeV for hadrons and muons. Electromagnetic particles below 10$^{-6}$ of the primary energy were statistically combined (\emph{thinned}) to optimize runtime~\cite{CORSIKA:2023jyz}. 
The energy threshold for transitioning from the low-energy model FLUKA~\cite{Ballarini:2024uxz} to Pythia~8/Angantyr is set at 100 GeV, whereas the CORSIKA~8 default threshold is $\sim$80 GeV when employing other high-energy models.

\subsection{Shower developement}

\begin{figure}[h]
  \noindent
  \begin{minipage}{0.49\textwidth}
    \centering
    \hspace*{-0.7cm}
    \includegraphics[width=1.1\linewidth]{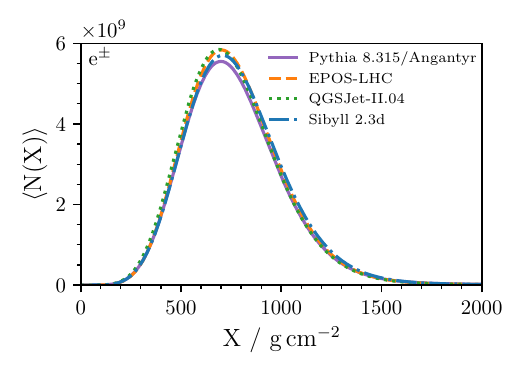}
  \end{minipage}%
  \hfill
  \begin{minipage}{0.49\textwidth}
    \centering
    \hspace*{-0.2cm}
    \includegraphics[width=1.1\linewidth]{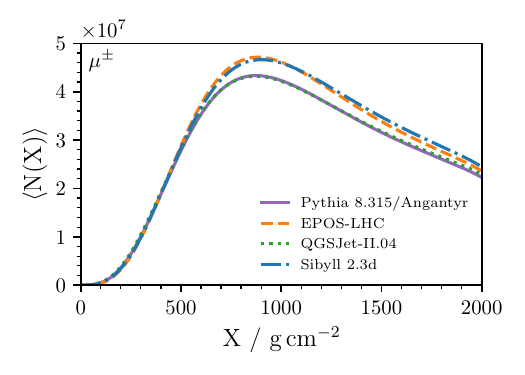}
  \end{minipage}
  \vspace*{-0.3cm}
  \captionof{figure}{%
  Average longitudinal shower profiles of electrons/positrons (left) and muons (right) for inclined ($\theta = 67^\circ$) iron-induced 10$^{19}$ eV air showers using FLUKA as low-energy interaction model, and Pythia 8.315/Angantyr (solid), EPOS-LHC (dashed), Sibyll 2.3d (dash-dotted) and QGSJet-II.04 (dotted) as high-energy interaction models.}
  \label{fig:long_e_mu}
\end{figure}

Air showers develop through a chain of interactions and energy losses through their passage in the atmosphere, leading to measurable observables, i.e. d$E$/d$X$, $X_\mathrm{\max}$. The number of identified particles at a given depth, from which the average longitudinal profile $\langle N(X) \rangle$ is derived, is shown in Fig.~\ref{fig:long_e_mu} for the mean e$^\pm$ and $\mu^\pm$ profiles as a function of atmospheric depth.

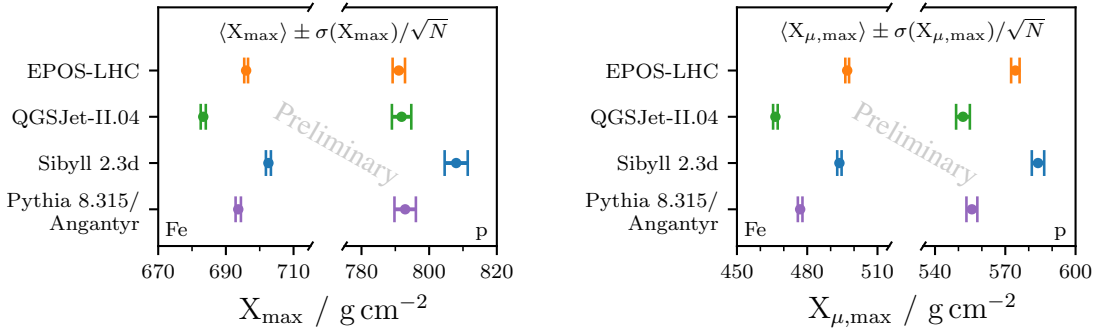
\begin{figure}[h]
  \noindent
  \begin{minipage}{0.49\textwidth}
    \centering
    \includegraphics[width=\linewidth]{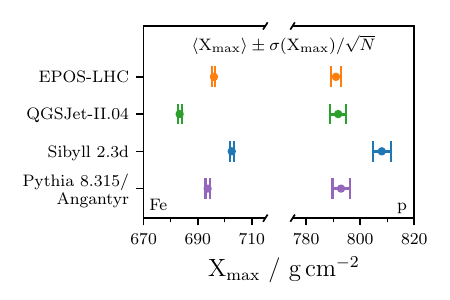}
    \begin{tikzpicture}[remember picture, overlay] 
      \node at (1, 3.2)[rotate=-30, text opacity=0.2] {Preliminary};
    \end{tikzpicture}
  \end{minipage}%
  \hfill
  \begin{minipage}{0.49\textwidth}
    \centering
    \includegraphics[width=\linewidth]{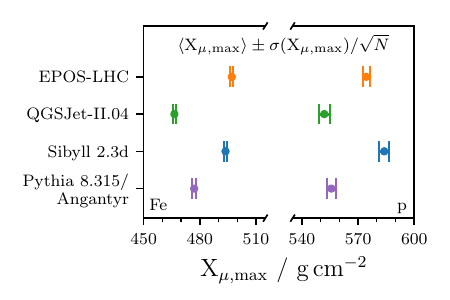}
    \begin{tikzpicture}[remember picture, overlay] 
      \node at (1, 3.2)[rotate=-30,text opacity=0.2] {Preliminary};
    \end{tikzpicture}
  \end{minipage}
  \vspace*{-0.5cm}
  \captionof{figure}{%
  Average depth of the shower maximum $X_\mathrm{\max}$ (left) and average depth of the muonic shower maximum $X_{\mu,\mathrm{\max}}$ for inclined ($\theta = 67^\circ$) iron-induced 10$^{19}$ eV air showers.}
  \label{fig:Xmax_Xmumax}
\end{figure}

In parallel, the energy deposit profile quantifies the energy released by the shower, mainly through bremsstrahlung, pair production, and particle decays, serving as a robust estimator of the shower size and its maximum depth $X_\mathrm{\max}$ displayed in Fig.~\ref{fig:Xmax_Xmumax} (left). The higher inelastic p-air cross-section predicted by Pythia8/Angantyr -- about 10\% larger than Sibyll2.3d, as shown in Fig.~\ref{fig:xs}, implies a shorter interaction length and thus a shallower first interaction, contributing to an increase in $X_{\max}$ relative to Sibyll 2.3d. Given its smaller proton-air cross section (Fig.~\ref{fig:xs}), QGSJet-II.04 would be expected to yield a deeper ${\langle X_\mathrm{\max}\rangle}_\mathrm{p}$, yet its values align with Pythia~8/Angantyr and EPOS-LHC. This indicates that differences in shower development beyond the first interaction, such as particle production mechanisms and secondary interactions, play a compensating role that masks the expected shift from the cross-section alone.

The muonic shower maximum, $X_{\mu,\mathrm{\max}}$, marks the depth in the atmosphere where muon production peaks. As illustrated in Fig.~\ref{fig:Xmax_Xmumax} (right), this observable is sensitive to the mass of the primary particle: proton-induced showers exhibit deeper $X_{\mu,\mathrm{\max}}$ values than iron-induced ones, reflecting a more extended hadronic cascade. For $X_{\mu,\mathrm{\max}}$, the model ordering is consistent between proton and iron primaries, reflecting intrinsic hadronic model features rather than primary mass. For protons, Sibyll~2.3d predicts the deepest values of both $X_{\max}$ and $X_{\mu,\mathrm{\max}}$, indicating common model features that delay both electromagnetic and hadronic cascades, pushing the shower maxima deeper.

\subsection{Muon puzzle and Pierre Auger measurement}

The introduction of Pythia~8/Angantyr into air shower physics is particularly relevant to gaining further insight into the \emph{Muon puzzle} -- a longstanding discrepancy between simulated and observed muons, such as those measured by the Pierre Auger Observatory. At a given observer level, the lateral distribution function describes the spatial spread of particles from the shower core, providing insight into the shower footprint. The energy spectra of charged particles at this level help characterize both the energy deposition and the particle composition, offering a deeper understanding of the shower's energy profile. The observer level is fixed at an altitude of 1400 m, approximately that of the Pierre Auger Observatory, and is not projected onto the shower plane. No corrections have been applied for asymmetry effects arising from this choice of observer plane. 

\begin{figure}[h]
  \noindent
  \begin{minipage}{0.49\textwidth}
    \centering
    \includegraphics[width=\linewidth]{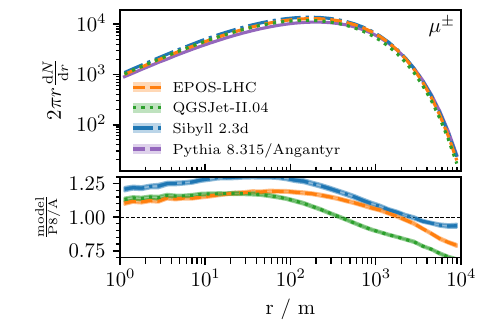}
    \label{fig:ldf_mu}
  \end{minipage}%
  \hfill
  \begin{minipage}{0.49\textwidth}
    \centering
    \includegraphics[width=\linewidth]{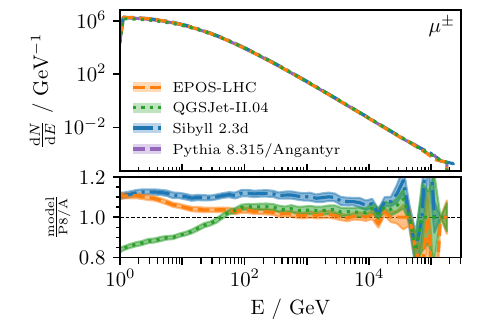}
    \label{fig:espect_mu} 
  \end{minipage}
  \vspace*{-1cm}
  \captionof{figure}{%
  Median lateral distributions (left) and median energy spectra (right) of muons at Auger height for inclined ($\theta = 67^\circ$) iron-induced 10$^{19}$ eV air showers. The shaded regions indicate the interquartile range (25\%-75\%).}
  \label{fig:ldf_espect_mu}
\end{figure}

Fig~\ref{fig:ldf_espect_mu} illustrates both the lateral distribution function and the energy spectrum of muons at this observation level. With up to 25\% fewer muons near the shower core than Sibyll 2.3d, and 10-15\% fewer than EPOS-LHC and QGSJet-II.04, Pythia~8/Angantyr appears less muon-rich at small distances, though it matches or exceeds the other models at larger distances.Pythia~8/Angantyr produces roughly 10\% fewer muons than Sibyll~2.3d in the 1 GeV to 10 TeV energy range. Similarly, when compared to EPOS-LHC, Pythia~8/Angantyr is slightly lower (up to 10\%) in the 1 to 100 GeV range but then matches at higher energies, while QGSJet-II.04 produces fewer low-energy muons (15\% less around 1 GeV) but rises above Pythia ($\sim5$\%) from 100 GeV until the highest muon energies where fluctuations dominate. Overall, these differences highlight Pythia~8/Angantyr as a promising target for further tuning efforts to better reproduce the muon content of air showers.

We apply the z-scale approach, originally developed for meta-analyses of muon density measurements across multiple experiments, to evaluate Pythia in the same context as in earlier works~\cite{Soldin:2021+r, ArteagaVelazquez:20236d}. The comparison between experimental data and Monte-Carlo simulations is carried out through the calculation of the z-scale, defined as 
\begin{equation}
  z = \frac{\ln{\langle N_\mu^\mathrm{\det} \rangle} - \ln{\langle N_{\mu,\mathrm{p}}^\mathrm{\det} \rangle}}{\ln{\langle N_{\mu,\mathrm{Fe}}^\mathrm{\det} \rangle} - \ln{\langle N_{\mu,\mathrm{p}}^\mathrm{\det} \rangle}}
\end{equation}
Here, $\langle N_\mu^\mathrm{\det} \rangle$ refers to the mean measured muon density, and $\langle N_{\mu,\mathrm{p}}^\mathrm{\det} \rangle$ ($\langle N_{\mu,\mathrm{Fe}}^\mathrm{\det} \rangle$) are the predicted average muon densities for proton (iron) cosmic-ray primaries. The z-scale is designed so that measurements consistent with iron primaries yield $z=1$, and those consistent with proton primaries yield $z=0$. 

\begin{figure}[h]
  \centering
  \includegraphics[width=0.5\linewidth]{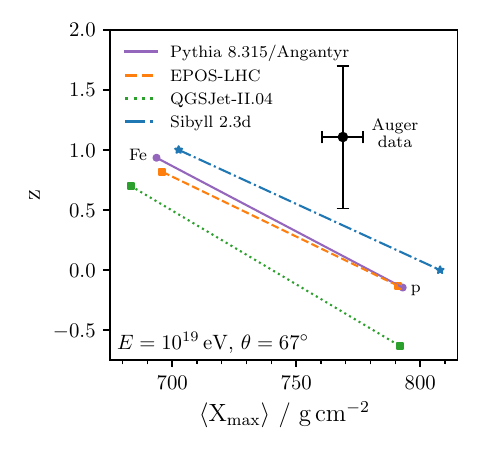}
  \begin{tikzpicture}[remember picture, overlay] 
    \node at (-3.2, 4)[rotate=45, text opacity=0.15] {\huge Preliminary};
  \end{tikzpicture}
  \caption{Muon density measurements from Auger~\cite{PierreAuger:2021qsd} converted to the z-scale using Sibyll 2.3d as reference.}
  \label{zscale}
\end{figure}
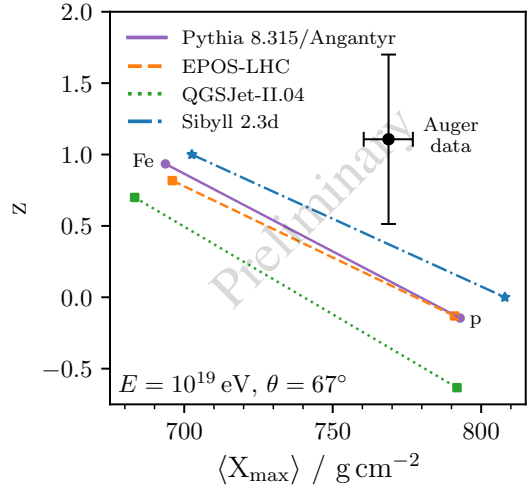

Simulations from Sibyll~2.3d from~\cite{PierreAuger:2021qsd} for the average logarithmic muon content $\langle  \ln R_\mu \rangle$, and Sibyll~2.3d within CORSIKA 8 for the number of muons $N_\mu$, as a function of the average shower depth $\langle X_\mathrm{\max}\rangle$, are used as reference to compute a z-scale frame to which both the Auger measurement and interaction model predictions are translated. For this approach to be valid, all simulations must be performed under identical conditions -- using the same primaries, primary energy, zenith angle, and particle tracking energy thresholds as the reference Sibyll~2.3d simulations. Fig.~\ref{zscale} displays this z-scale frame as a function of the average shower depth $\langle X_\mathrm{\max} \rangle$ where the compatibility of interaction models with Auger measurement can be discussed. 

As shown in Fig.~\ref{zscale}, in its default configuration, Pythia~8.315/Angantyr underestimates the muon content observed by Auger, highlighting its potential for further tuning efforts.

Pythia~8/Angantyr and EPOS-LHC are close to each other in terms of z-scale for both proton and iron primaries, with Pythia slightly higher for iron. This can stems from the limited deviations of EPOS-LHC in the muon energy spectra, while the muon lateral distribution remains within $\sim10$\% of Pythia. The observed differences between QGSJet-II.04 and Pythia from Fig~\ref{fig:ldf_espect_mu} likely underlie the gap in z-scale between the two models, with QGSJet exhibiting a lower z-scale than Pythia and consequently deviating further from Auger measurements. By contrast, Pythia~8/Angantyr yields less muon-rich and slightly shallower showers than Sibyll~2.3d for both proton and iron primaries, explaining the larger z-scale difference observed between these two models. 

\section*{Conclusion}

We presented a comparison of particle production between Pythia~8/Angantyr and other interaction models. Earlier results~\cite{Reininghaus:2023ctx} are outdated, as the present predictions are obtained with a significantly improved model. While Pythia reproduces key features, such as the rise of multiplicity with energy and overall energy flow, it requires tuning to match data more precisely. The first full air shower simulations with Pythia~8/Angantyr for iron primaries were also performed, providing insight into its behavior in extensive air shower development.

Overall, Pythia~8/Angantyr produces ${\langle X_\mathrm{\max}\rangle}$ values that lie within the range spanned by the other models, occupying a region similar to that of EPOS-LHC and between the extremes of Sibyll 2.3d and QGSJet-II.04. For $X_{\mu,\mathrm{\max}}$, Pythia predicts a shallower muonic shower maximum compared to EPOS-LHC and Sibyll, though not as shallow as QGSJet. For the muon lateral distribution at Auger altitude, Pythia~8/Angantyr predicts fewer muons near the shower core compared to the other models, while matching or exceeding the other models at larger distances. In terms of the muon energy spectrum at ground, Pythia yields more low-energy muons than QGSJet, but fewer than EPOS-LHC and Sibyll; at energies above $\sim100$ GeV, it predicts fewer high-energy muons than QGSJet and Sibyll, while being comparable to EPOS-LHC.

Finally, when comparing to the muon content measured by the Pierre Auger Observatory, we find that Pythia~8/Angantyr is not compatible with Auger measurements, as it predicts shallower $X_\mathrm{\max}$ and lower muon content.
The next step is to tune Pythia to improve its agreement with proton-proton and fixed-target measurements, and investigate the resulting impact on air shower predictions and compatibility with Auger data~\cite{Gaudu:2024mkp,Windau:ICRC2025}.

%\bibliographystyle{JHEP}
%\bibliography{skeleton}

%\let\oldbibliography\thebibliography
%\renewcommand{\thebibliography}[1]{%
%  \oldbibliography{#1}%
%  \setlength{\itemsep}{1pt}%
%}

%{\footnotesize
%\bibliographystyle{JHEPnotitle}
%\bibliography{skeleton}

\providecommand{\href}[2]{#2}\begingroup\raggedright\begin{thebibliography}{10}

\bibitem{Engel:2018akg}
R.~Engel, D.~Heck, T.~Huege, T.~Pierog, M.~Reininghaus, F.~Riehn
  et~al.\href{https://doi.org/10.1007/s41781-018-0013-0}{\emph{Comput. Softw.
  Big Sci.} {\bfseries 3} (2019) 2}
  [\href{https://arxiv.org/abs/1808.08226}{{\ttfamily 1808.08226}}].

\bibitem{Alameddine:2024cyd}
J.M.~Alameddine
  et~al.\href{https://doi.org/10.1016/j.astropartphys.2024.103072}{\emph{Astropart.
  Phys.} {\bfseries 166} (2025) 103072}
  [\href{https://arxiv.org/abs/2409.15999}{{\ttfamily 2409.15999}}].

\bibitem{Pierog:2013ria}
T.~Pierog, I.~Karpenko, J.M.~Katzy, E.~Yatsenko and
  K.~Werner\href{https://doi.org/10.1103/PhysRevC.92.034906}{\emph{Phys. Rev.
  C} {\bfseries 92} (2015) 034906}
  [\href{https://arxiv.org/abs/1306.0121}{{\ttfamily 1306.0121}}].

\bibitem{Ostapchenko:2010vb}
S.~Ostapchenko\href{https://doi.org/10.1103/PhysRevD.83.014018}{\emph{Phys.
  Rev. D} {\bfseries 83} (2011) 014018}
  [\href{https://arxiv.org/abs/1010.1869}{{\ttfamily 1010.1869}}].

\bibitem{Riehn:2019jet}
F.~Riehn, R.~Engel, A.~Fedynitch, T.K.~Gaisser and
  T.~Stanev\href{https://doi.org/10.1103/PhysRevD.102.063002}{\emph{Phys. Rev.
  D} {\bfseries 102} (2020) 063002}
  [\href{https://arxiv.org/abs/1912.03300}{{\ttfamily 1912.03300}}].

\bibitem{Gaudu:2024rsq}
{\scshape CORSIKA 8} collaboration, \emph{{CORSIKA 8 with Pythia 8/Angantyr:
  Simulating Vertical Proton Showers}},  in \emph{{28th European Cosmic Ray
  Symposium}}, 12, 2024 [\href{https://arxiv.org/abs/2412.15094}{{\ttfamily
  2412.15094}}].

\bibitem{Gaudu:2025K1}
C.~Gaudu, M.~Reininghaus and F.~Riehn, \emph{{CORSIKA 8 with Pythia 8/Angantyr:
  Simulating Inclined Proton Showers}},  in \emph{Proceedings of 7th
  International Symposium on Ultra High Energy Cosmic Rays {\textemdash}
  PoS(UHECR2024)}, vol.~484, p.~089, 2025,
  \href{https://doi.org/10.22323/1.484.0089}{DOI}.

\bibitem{Bierlich:2022pfr}
C.~Bierlich
  et~al.\href{https://doi.org/10.21468/SciPostPhysCodeb.8}{\emph{SciPost Phys.
  Codeb.} {\bfseries 2022} (2022) 8}
  [\href{https://arxiv.org/abs/2203.11601}{{\ttfamily 2203.11601}}].

\bibitem{Bierlich:2018xfw}
C.~Bierlich, G.~Gustafson, L.~L\"onnblad and
  H.~Shah\href{https://doi.org/10.1007/JHEP10(2018)134}{\emph{JHEP} {\bfseries
  10} (2018) 134} [\href{https://arxiv.org/abs/1806.10820}{{\ttfamily
  1806.10820}}].

\bibitem{Bierlich:2021poz}
C.~Bierlich, T.~Sj\"ostrand and
  M.~Utheim\href{https://doi.org/10.1140/epja/s10050-021-00543-3}{\emph{Eur.
  Phys. J. A} {\bfseries 57} (2021) 227}
  [\href{https://arxiv.org/abs/2103.09665}{{\ttfamily 2103.09665}}].

\bibitem{Andersson:1992iq}
B.~Andersson, G.~Gustafson and
  H.~Pi\href{https://doi.org/10.1007/BF01474343}{\emph{Z. Phys. C} {\bfseries
  57} (1993) 485}.

\bibitem{Sjostrand:2021dal}
T.~Sj\"ostrand and
  M.~Utheim\href{https://doi.org/10.1140/epjc/s10052-021-09953-5}{\emph{Eur.
  Phys. J. C} {\bfseries 82} (2022) 21}
  [\href{https://arxiv.org/abs/2108.03481}{{\ttfamily 2108.03481}}].

\bibitem{PierreAuger:2012egl}
{\scshape Pierre Auger}
  collaboration\href{https://doi.org/10.1103/PhysRevLett.109.062002}{\emph{Phys.
  Rev. Lett.} {\bfseries 109} (2012) 062002}
  [\href{https://arxiv.org/abs/1208.1520}{{\ttfamily 1208.1520}}].

\bibitem{ALICE:2015qqj}
{\scshape ALICE}
  collaboration\href{https://doi.org/10.1016/j.physletb.2015.12.030}{\emph{Phys.
  Lett. B} {\bfseries 753} (2016) 319}
  [\href{https://arxiv.org/abs/1509.08734}{{\ttfamily 1509.08734}}].

\bibitem{ALICE:2015olq}
{\scshape ALICE}
  collaboration\href{https://doi.org/10.1140/epjc/s10052-016-4571-1}{\emph{Eur.
  Phys. J. C} {\bfseries 77} (2017) 33}
  [\href{https://arxiv.org/abs/1509.07541}{{\ttfamily 1509.07541}}].

\bibitem{ALICE:2010cin}
{\scshape ALICE}
  collaboration\href{https://doi.org/10.1140/epjc/s10052-010-1339-x}{\emph{Eur.
  Phys. J. C} {\bfseries 68} (2010) 89}
  [\href{https://arxiv.org/abs/1004.3034}{{\ttfamily 1004.3034}}].

\bibitem{UA5:1986yef}
{\scshape UA5} collaboration\href{https://doi.org/10.1007/BF01410446}{\emph{Z.
  Phys. C} {\bfseries 33} (1986) 1}.

\bibitem{Basu:2020jbk}
S.~Basu, S.~Thakur, T.K.~Nayak and
  C.A.~Pruneau\href{https://doi.org/10.1088/1361-6471/abc05c}{\emph{J. Phys. G}
  {\bfseries 48} (2020) 025103}
  [\href{https://arxiv.org/abs/2008.07802}{{\ttfamily 2008.07802}}].

\bibitem{Gaudu:2024mkp}
C.~Gaudu, \emph{{Pythia 8 and Air Shower Simulations: A Tuning Perspective}},
  in \emph{{22nd International Symposium on Very High Energy Cosmic Ray
  Interactions}}, 10, 2024 [\href{https://arxiv.org/abs/2411.00111}{{\ttfamily
  2411.00111}}].

\bibitem{Reininghaus:2022gnr}
{\scshape CORSIKA 8}
  collaboration\href{https://doi.org/10.21468/SciPostPhysProc.15.019}{\emph{SciPost
  Phys. Proc.} {\bfseries 15} (2024) 019}
  [\href{https://arxiv.org/abs/2210.07797}{{\ttfamily 2210.07797}}].

\bibitem{Reininghaus:2023ctx}
{\scshape CORSIKA 8}
  collaboration\href{https://doi.org/10.1051/epjconf/202328305010}{\emph{EPJ
  Web Conf.} {\bfseries 283} (2023) 05010}
  [\href{https://arxiv.org/abs/2303.02792}{{\ttfamily 2303.02792}}].

\bibitem{Ballarini:2024uxz}
F.~Ballarini et~al.\href{https://doi.org/10.1051/epjn/2024015}{\emph{EPJ
  Nuclear Sci. Technol.} {\bfseries 10} (2024) }.

\bibitem{Soldin:2021+r}
D.~Soldin, \emph{{Update on the Combined Analysis of Muon Measurements from
  Nine Air Shower Experiments}},  in \emph{Proceedings of 37th International
  Cosmic Ray Conference {\textemdash} PoS(ICRC2021)}, vol.~395, p.~349, 2021,
  \href{https://doi.org/10.22323/1.395.0349}{DOI}.

\bibitem{ArteagaVelazquez:20236d}
J.C.~Arteaga~Velazquez, \emph{{A report by the WHISP working group on the
  combined analysis of muon data at cosmic-ray energies above 1 PeV}},  in
  \emph{Proceedings of 38th International Cosmic Ray Conference {\textemdash}
  PoS(ICRC2023)}, vol.~444, p.~466, 2023,
  \href{https://doi.org/10.22323/1.444.0466}{DOI}.

\bibitem{PierreAuger:2021qsd}
{\scshape Pierre Auger}
  collaboration\href{https://doi.org/10.1103/PhysRevLett.126.152002}{\emph{Phys.
  Rev. Lett.} {\bfseries 126} (2021) 152002}
  [\href{https://arxiv.org/abs/2102.07797}{{\ttfamily 2102.07797}}].

\end{thebibliography}\endgroup
%}

\setlength{\bibsep}{0pt}
{\footnotesize
\begin{multicols}{2}

\end{multicols}
}

\newpage
\clearpage
\section*{The CORSIKA 8 Collaboration}

\begin{sloppypar}\noindent
J.M.~Alameddine$^{1,2}$,
J.~Albrecht$^{1,2}$,
A.A.~Alves Jr.$^{3,4}$,
J.~Ammerman-Yebra$^{5}$,
L.~Arrabito$^{6}$,
D.~Baack$^{1,2}$,
R.~Cesista$^{6}$,
A.~Coleman$^{7}$,
C.~Deaconu$^{8}$,
H.~Dembinski$^{1,2}$,
D.~Elsässer$^{1,2}$,
R.~Engel$^{3}$,
A.~Faure$^{6}$,
A.~Ferrari$^{3}$,
C.~Gaudu$^{9}$,
C.~Glaser$^{7,1}$,
M.~Gottowik$^{3}$,
D.~Heck$^{3}$,
T.~Huege$^{3,10}$,
K.H.~Kampert$^{9}$,
N.~Karastathis$^{3}$,
L.~Nellen$^{11}$,
D.~Parello$^{12,13}$,
T.~Pierog$^{3}$,
R.~Prechelt$^{14}$,
M.~Reininghaus$^{15}$,
W.~Rhode$^{1,2}$,
F.~Riehn$^{1}$,
M.~Sackel$^{1,2}$,
P.~Sampathkumar$^{3}$,
A.~Sandrock$^{9}$,
J.~Soedingrekso$^{1,2}$,
R.~Ulrich$^{3}$,
P.~Windischhofer$^{8}$,
B.~Yue$^{9}$

\vspace{1ex}
\begin{center}
\rule{0.1\columnwidth}{0.5pt}
\raisebox{-0.4ex}{\scriptsize$\bullet$}
\rule{0.1\columnwidth}{0.5pt}
\end{center}
\vspace{1ex}

\begin{itemize}[labelsep=0.2em,align=right,labelwidth=0.7em,labelindent=0em,leftmargin=2em,noitemsep,before={\renewcommand\makelabel[1]{##1 }}]
\item[$^{1}$] Technische Universität Dortmund (TU), Department of Physics, Dortmund, Germany
\item[$^{2}$] Lamarr Institute for Machine Learning and Artificial Intelligence, Dortmund, Germany
\item[$^{3}$] Karlsruhe Institute of Technology (KIT), Institute for Astroparticle Physics (IAP), Karlsruhe, Germany
\item[$^{4}$] University of Cincinnati, Cincinnati, OH, United States
\item[$^{5}$] IMAPP, Radboud University Nijmegen, Nijmegen, The Netherlands
\item[$^{6}$] Laboratoire Univers \& Particules de Montpellier, CNRS \& Université de Montpellier (UMR-5299), 34095 Montpellier, France
\item[$^{7}$] Uppsala University, Department of Physics and Astronomy, Uppsala, Sweden
\item[$^{8}$] Department of Physics, Enrico Fermi Institute, Kavli Institute for Cosmological Physics, University of Chicago, Chicago, IL 60637, USA
\item[$^{9}$] Bergische Universität Wuppertal, Department of Physics, Wuppertal, Germany
\item[$^{10}$] Vrije Universiteit Brussel, Astrophysical Institute, Brussels, Belgium
\item[$^{11}$] Universidad Nacional Autónoma de México (UNAM), Instituto de Ciencias Nucleares, México, México
\item[$^{12}$] DALI, Univ Perpignan, Perpignan, France
\item[$^{13}$] LIRMM Univ Montpellier, CNRS, Montpellier, France
\item[$^{14}$] University of Hawai'i at Manoa, Department of Physics and Astronomy, Honolulu, USA
\item[$^{15}$] Independent researcher
\end{itemize}

\end{sloppypar}

\section*{Acknowledgements}
\noindent
We thank T.~Sjöstrand, L.~Lönnblad, and the Pythia~8 collaborators for their support in the implementation of Pythia~8/Angantyr in CORSIKA~8.

This research was funded by the Deutsche Forschungsgemeinschaft (DFG, German Research Foundation) – Projektnummer 445154105 and Collaborative Research Center SFB1491 "Cosmic Interacting Matters - From Source to Signal". This research has been partially funded by the Federal Ministry of Education and research of Germany and the state of North Rhine-Westphalia as part of the Lamarr Institute for Machine Learning and Artificial Intelligence. We acknowledge support through project UNAM-PAPIIT IN114924.

This work has also received financial support from Ministerio de Ciencia e Innovación/Agencia Estatal de Investigación (PRE2020-092276). A.~Coleman is supported by the Swedish Research Council (Vetenskapsrådet) under project no. 2021-05449. C.~Glaser is supported by the Swedish Research Council (Vetenskapsrådet) under project no. 2021-05449 and the European Union. F.~Riehn received funding from the European Union’s Horizon 2020 research and innovation programme under the Marie Skłodowska-Curie grant agreement No.~101065027. P.~Windischhofer and C.~Deaconu thank the NSF for Award 2411662. The authors acknowledge support by the High Performance and Cloud Computing Group at the Zentrum für Datenverarbeitung of the University of Tübingen, the state of Baden-Württemberg through bwHPC and the German Research Foundation (DFG) through grant no INST 37/935-1 FUGG. The computations were partially carried out on the PLEIADES cluster at the University of Wuppertal, which was supported by the Deutsche Forschungsgemeinschaft (DFG, grant No. INST 218/78-1 FUGG) and the Bundesministerium für Bildung und Forschung (BMBF).

\end{document}